\documentclass[aps,prl,twocolumn,showpacs,showkeys,preprintnumbers]{revtex4}
\usepackage{epsfig}
\newcommand{\beq}{\begin{equation}}
\newcommand{\eeq}{\end{equation}}
\newcommand{\beeq}{\begin{eqnarray}}
\newcommand{\eeeq}{\end{eqnarray}}
\newcommand{\nn}{\nonumber}
\newcommand\as{{\alpha_s}}

\begin{document}
\preprint{IPPP/01/48, DCPT/01/96, hep-ph/0110315}
\title{Three-jet cross sections in hadron-hadron collisions at next-to-leading order}
\author{Zolt\'an Nagy}
\email[]{Zoltan.Nagy@durham.ac.uk}
\homepage[]{http://www.cpt.dur.ac.uk/~nagyz/nlo++.html}
\affiliation{Department of Physics, University of Durham DH1 3LE, England}
\date{\today}

\begin{abstract}
We present a new QCD event generator for hadron collider which can calculate 
one-, two- and three-jet cross sections at next-to-leading order accuracy.
In this letter we study the transverse energy spectrum of three-jet 
hadronic events using the $k_\perp$ algorithm. We show that the next-to-leading
order correction significantly reduces the renormalization and factorization 
scale dependence of the three-jet cross section.
\end{abstract}

\maketitle

The latest version of the experiment at Tevatron and the future collider 
LHC will provide precise data so that not only inclusive measurements can 
be used to study the physics of hadronic final state. The studies of the 
event shapes and multi-jet event can be important projects.   

One of the important theoretical tools in the analysis of hadronic
final states is perturbative Quantum Chromodynamics (QCD).  In order to
make quantitative predictions in perturbative QCD, it is essential to
perform the computations (at least) at the next-to-leading order (NLO)
accuracy. In hadron collision the most easily calculated one- and two-jet
cross sections have so far been calculated at NLO level \cite{jetrad, ks}.
At next-to-leading level some three-jet observables were calculated by 
Giele and Kilgore \cite{Kilgore:2000dr,Kilgore:1997sq} and by Tr\'ocs\'anyi 
\cite{Trocsanyi:1996by}.  
In this letter we present a new NLO event generator for calculating jet 
observables in hadron-hadron collision. We compute the three-jet cross 
sections using the $k_\perp$ algorithm \cite{kT-alg, kT-soper} to resolve 
jets in the final state. With our Monte Carlo program one can 
compute the NLO cross section for any other infrared safe one-, two- and 
three-jet observables. The presented distributions are given simply as 
illustration. 

In the case of one-jet inclusive cross section in high $p_T$ region the 
forthcoming experimental data requires the knowledge of the higher order 
corrections. Recently, progress has been made in calculating the two loop 
$2\to 2$ matrix elements \cite{Anastasiou:2000ue,Anastasiou:2001sv,Glover:2001rd}. 
These matrix element are need to set up a Monte Carlo program for calculating 
the next-to-next-to-leading order (NNLO) one- and two-jet cross section. 
However, the three-jet NLO calculation, comprising the one-loop $2\to 3$ and
tree-level $2\to 4$ processes is a necessary first step of this project. A 
numerically stable and fast three-jet NLO program can provide enough hope 
that it might be possible to develop a numerically stable Monte Carlo program 
for calculating one- and two-jet cross sections at NNLO level.

In the last few years the theoretical developments make possible the 
next-to-leading order calculation for the three-jet quantities. There are 
several general methods available for the cancellation of the infrared 
divergences that can be used for setting up a Monte Carlo program 
\cite{Nagy:1997bz,Frixione:1996ms,Catani:1997vz}. In computing the NLO correction we use the dipole 
formalism of Catani-Seymour \cite{Catani:1997vz} that we modified slightly 
in order to have a better control on the numerical calculation. The main idea
is to cut the phase space of the dipole subtraction terms as introduced in 
Ref. \cite{Nagy:1999bb}, the details of how to this to apply for the case of 
hadron-hadron scattering will be given elsewhere.    

The advantages of using the dipole method  are the followings: i) no
approximation is made; ii) the exact phase space factorization allows
full control over the efficient generation of the phase space; iii)
neither the use of color ordered subamplitudes, nor symmetrization, nor
partial fractioning of the matrix elements is required; iv) Lorentz
invariance is maintained, therefore, the switch between various frames can 
be achieved by simply transforming the momenta; v) the use of crossing 
functions is avoided; vi) it can be implemented in an actual program in a 
fully process independent way.

In this calculation we used the crossing symmetric tree- and one-loop level
amplitudes. The parton subprocess $0\to ggggg$ \cite{amp-g5}, 
$0\to q\bar{q}ggg$ \cite{amp-q2g3}, $0\to q\bar{q}Q\bar{Q}g$ \cite{amp-q4g1} 
and the subprocesses related to these have been computed to one-loop and  
$0\to gggggg$, $0\to q\bar{q}gggg$, $0\to q\bar{q}Q\bar{Q}gg$ 
\cite{Gunion:1985bp,Gunion:1986zh,Gunion:1986zg,amp-six}
and the crossed processes have been computed at tree level. 

We have checked numerically that in all soft and collinear regions the
difference of the real and subtraction terms contain only integrable
square-root singularities. Furthermore, we have also checked that our
results are independent of the parameter that controls the volume of
the cut dipole phase space, which ensures that indeed the same quantity
has been subtracted from the real correction as added to the virtual
one.

Finally, to have a further check of the computation, 
Our method of implementing the dipole substraction terms allows for the 
construction of a process independent programming of QCD jet cross sections 
at the NLO accuracy.  We use the same program structure, with trivial 
modifications, to compute one-, two- and three-jet cross sections. 

\begin{figure}
\centering
\epsfig{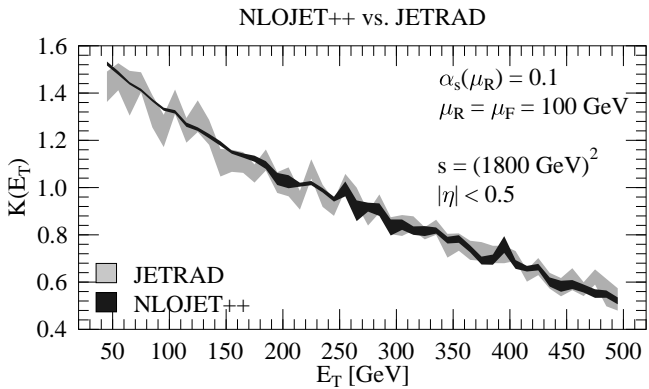}
\caption{\label{fig:jetrad-nlojet} Comparison of the $K$ factors of the 
  one-jet inclusive cross section defined using the $k_\perp$ and for 
  $\rm MRSD'_{-}$ parton densities obtained with Monte Carlo programs 
  JETRAD and NLOJET++ (this work). The bands indicate the statistical 
  error of the calculations.}
\end{figure}

In order to check the base structure of the program we compare the our 
inclusive one-jet NLO result to prediction of the program {\tt JETRAD} 
\cite{jetrad}. 
In this comparison we compare the one-jet inclusive cross section 
using the $k_\perp$ jet algorithm and $\rm MRSD_-'$ parton distribution 
function \cite{mrsd-p}. We find good agreement between the two program.
The differences are within the statistical error as 
Fig. \ref{fig:jetrad-nlojet} shows.

Historically, only the cone algorithm has been used to reconstruct the jet 
at hadron collider. In the three or higher jet calculation at NLO level the 
cone algorithm is not suitable because it has a lot of difficulties: an 
arbitrary procedure must be implemented to split and merge the overlapping 
cones, and an ad-hoc parameter $R_{sep}$ is required to accomodate the 
difference between the jet definitions at parton and detector level. To avoid 
this uncertainty we use the $k_T$ algorithm which has been developed by several 
groups \cite{kT-alg,kT-soper}. Our implementation is based on the 
Ref. \cite{kT-soper}. The algorithm starts with a list of the particles and the 
empty list of the jets.
\begin{enumerate}
\item For each particle (pseudo-particle) $i$ in the list, define 
  \beq
  d_i = p^2_{T,i}\;\;.
  \eeq
For each pair $(i,j)$ of momenta $(i\neq j)$, define
  \beq
  \label{dij}
  d_{ij} = \min(p^2_{T,i}, p^2_{T,j})
  \frac{\Delta R_{ij}^2}{D^2}\;\;,
  \eeq
  where $\Delta R^2_{ij} = (\eta_i-\eta_j)^2+(\phi_i-\phi_j)^2$ is square of the 
angular separation which is expressed in the term of the pseudo-rapidity  
$\eta_i$ and the azimuth angle $\phi_i$. $D$ is a free parameter. 
The usual choice of this parameter is $D=1$.
\item Find the minimum of all the $d_i$ and $d_{ij}$ and label it $d_{\mathrm{min}}$.  
\item If $d_{\mathrm{min}}$ is $d_{ij}$, remove particles (pseudo-particles) $i$ and $j$
  from the list and replace them with a new, merged pseudo-particle $p_{(ij)}$ 
  given by the recombination scheme. In this paper we use the $E$ 
  recombination scheme which is define the new pseudo-particle as the sum of 
  the two particle 
  \beq
  p_{(ij)} = p_i+p_j\;\;.
  \eeq
\item If $d_{\mathrm{min}}$ is $d_i$, remove particle (pseudo-particle) $i$ 
  from the list of particle and add it to list of jets.
\item If any particles remain, go to step 1.
\end{enumerate} 
The algorithm produces a list of jets, each separated by $\Delta R_{ij} > D$. 

Once the phase space integrations are carried out, we write the NLO jet cross 
section in the following form:
\beeq\nn
\label{signjet}
\sigma^{n\rm{jet}}_{AB} &=& \sum_{a,b} \int d\eta_ad\eta_b
f_{a/A}(\eta_a, \mu_F^2) f_{b/B}(\eta_b, \mu_F^2) \\
&\times&\hat{\sigma}_{ab}^{n\rm{jet}}
\left[p_a, p_b,\as(\mu_R^2), \mu_R^2/Q_{H}^2, \mu_F^2/Q_{H}^2\right]\;\;,  
\eeeq 
where $f_{i/H}(\eta, \mu_F^2)$ represents the patron distribution function of the 
incoming hadron defined at the factorization scale $\mu_F = x_F Q_{H}$,
$\eta_{a,b}$ is the fraction of the proton momentum carried by the scattered 
partons $p_{a,b}$, $Q_{H}$ is the hard scale that characterizes the parton 
scattering which could be $E_T$ of the jet, jet mass of the event, {\it etc}
and $\mu_R = x_R Q_{H}$ is the renormalization scale.
\begin{figure}
\centering
\epsfig{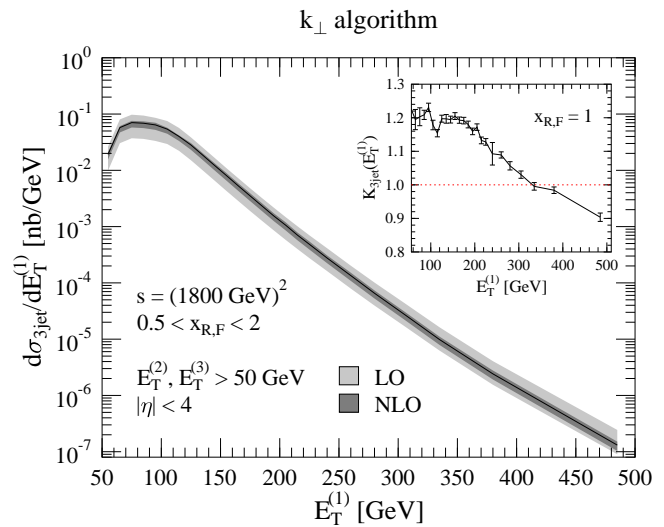}
\caption{\label{fig:sig-et1} The perturbative prediction for the three-jet 
  differential cross section in the term of the transverse energy of the 
  leading jet at Born level (light gray band) and next-to-leading order 
  (dark gray band). The bands indicate the theoretical uncertainty due to the 
  variation of the renormalization and factorization scales $x_{R,F}$ between 
  $0.5$ and $2$. The solid line is the NLO result for the $x_R = x_F = 1$ 
  choice of the scales. 
  }
\end{figure}

Eq.~(\ref{signjet}) shows that using the dipole method one may either
compute the full cross section at NLO accuracy including the
convolution with the parton distribution functions, or simply the 
parton level cross section $\hat{\sigma}_{ab}$
which can then be convoluted with the parton densities after the Monte
Carlo integration. The latter procedure is the proper one if we are
interested in measurement of the parton distribution functions (to avoid the
recalculation of the Monte Carlo integrals after each step of the
fitting iteration). 

The three-jet cross sections presented here were calculated for 
TEVATRON collider in proton-antiproton collision at the center of mass energy 
$\sqrt{s} = 1800$ GeV. We restrict the pseudo-rapidity range and the minimum 
transverse energy of the jets in laboratory frame to be
\beq
-4 < \eta_{\mathrm{jet}} < 4\;\;,\qquad\;\; E_T > 50\ \mathrm{GeV}\;\;.
\eeq
We choose the transverse energy of the leading jets
\beq
Q_H = E_T^{(1)}\;\;,
\eeq
as the hard scattering scale.
\begin{figure}
  \epsfig{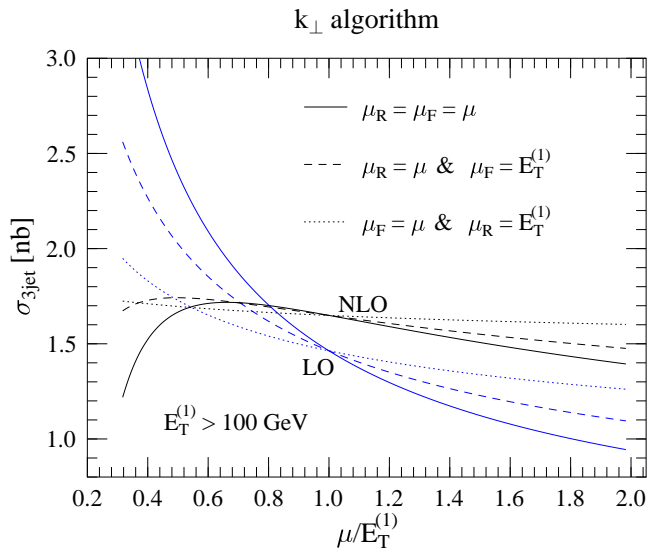}
\caption{\label{fig:mudep} The dependence of the three-jet cross section 
  $\sigma_{\rm 3jet}$ on the renormalization and factorization scales. }
\end{figure}

In Fig.~\ref{fig:sig-et1}, we plotted differential cross section in the term 
of the the transverse energy of leading jet convoluted with the CTEQ5M1 parton
distribution functions \cite{CTEQ5} and using the two-loop formula for the strong coupling, 
\beeq
\label{twoloopas}
&&\as(\mu) = \frac{\as(M_Z)}{w(\mu)}
\left(1-\frac{\as(M_Z)}{2\pi}\,\frac{\beta_1}{\beta_0} \frac{\ln(w(\mu))}{w(\mu)}\right)\;\;,\qquad\\
&&w(\mu) = 1 - \beta_0\,\frac{\as(M_Z)}{2\pi}\,\ln\left(\frac{M_Z}{\mu}\right)\;\;,
\eeeq
where $\as(M_Z) = 0.118$ and $\beta_0 = (11C_A - 4T_RN_f)/3$, 
$\beta_1 = (17C_A^2 - 6 C_F T_R N_f - 10 C_A T_R N_f)/3$, with $N_f = 5$ 
flavors. For the leading order results we used the
CTEQ5L distributions and the one-loop $\as$ ($\as(M_Z) = 0.127$ and 
$\beta_1 = 0$ in Eq.~(\ref{twoloopas})).
In this figure the theoretical uncertainty of the three-jet cross section 
is shown. Over the wide range of the value the renormalization and 
factorization scale ($0.5 < x_{R,F} < 2$) this uncertainty is 
in the next-to-leading order result is much than in the Born level calculation.

In Fig.~\ref{fig:mudep} we study the scale dependences of the three-jet
cross section. The strong dependence on the renormalization scale observed 
at LO is significantly reduced. The factorization scale dependence is not
significant at NLO and does not change much. Setting the two scales
equal, $\mu_R = \mu_F = \mu$, we can observe a wide plato peaking around
$x_R = x_F = 0.7$.
\begin{figure}
  \epsfig{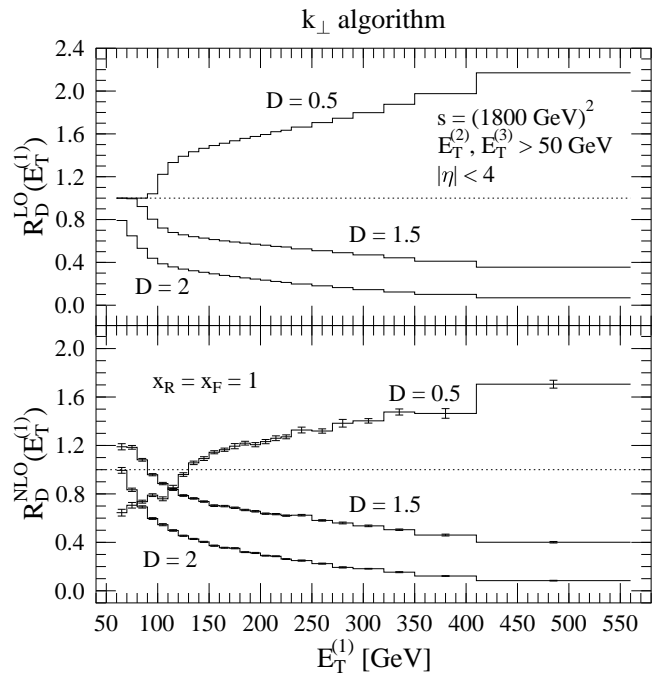}
\caption{\label{fig:Ddep} The dependence of the three-jet differential 
  cross section on the parameter $D$. The $R_D$ means the ration of the 
  differential cross sections for a given $D$ and for $D = 1$. Upper 
  figure shows the Born level result and the lower figure shows the NLO 
  prediction. The error bars indicate the statistical error of the Monte 
  Carlo calculation.}
\end{figure}

The inset figure in Fig.~\ref{fig:sig-et1} shows the $K$ factor (ratio of the
three-jet cross section at NLO to that at LO accuracy),
indicating the relative size of the correction. We can see the size of 
the NLO correction is between $10$ and $25\%$ for smaller values of the 
transverse energy and at the end of the spectra the size of the correction 
is almost zero. The error bars indicate the statistical error of the Monte 
Carlo calculation. Because of the strong logarithmic behavior of the cross 
section the Monte Carlo calculation is very sensitive to the 
``missed binning'' (when a huge positive comes from the real term and the 
corresponding huge negative weight form the subtraction term are filled in 
different histogram bins).  

In Fig.~\ref{fig:Ddep} we study the dependence of the differential cross 
section on parameter $D$. We plotted the ratios of the cross section 
\beq
R_D(E_T^{(1)}) = \frac{d\sigma^{3jet}_{p\bar{p}}(E_T^{(1)}; D)}{dE_T^{(1)}}
\left/\frac{d\sigma^{3jet}_{p\bar{p}}(E_T^{(1)};1)}{dE_T^{(1)}}\right.\;\;,
\eeq
 for three different values of the parameter $D$ ($D = 0.5,\, 1.5,\, 2$).  
The parameter $D$ controls the angular separation in the jet algorithm 
procedure in Eq.~\ref{dij}. Changing this parameter we expect more resolved 
jets (more three-jet events) with high transverse energy and less 
recombination for smaller values of $D$ and vice-versa. This behavior can 
be clearly observed from in Fig.~\ref{fig:Ddep}.

In this letter we presented a NLO computation of the three-jet cross 
section defined with the $k_\perp$ clustering algorithm in in hadron-hadron 
collision.  Our results were obtained using a partonic Monte Carlo program 
that is suitable for implementing any detector cuts. We found that the 
NLO the correction is under $30\%$ in the case of differential cross section
but the $K$ factor is sensitive to the allowed kinematic region. We 
demonstrated that the NLO corrections reduce the scale dependence 
significantly. We also presented how the differential cross section depends on 
the angular separation parameter $D$ used to define the jet. The same program can be used for 
computing the QCD radiative corrections to the (differential) cross 
section of any kind of one-, two-, or three-jet cross section or event-shape 
distribution in hadron-hadron collision. We compared the two-jet results 
obtained by our program to previous results and found agreement.

I thank Nigel Glover, Adrian Signer and Zol\-t\'an Tr\'o\-cs\'a\-nyi for 
the helpful discussions. This work was supported in part by the EU Fourth 
Framework Programme ``Training and Mobility of Researchers'', Network 
``QCD and particle structure'', contract FMRX-CT98-0194 (DG 12 - MIHT), 
the EU Fifth Framework Programme `Improving Human Potential', Research 
Training Network `Particle Physics Phenomenology at High Energy Colliders', 
contract HPRN-CT-2000-00149 as well as by the Hungarian Scientific Research 
Fund grant OTKA T-025482.


\bibliography{pp3jet}
\end{document}